# The fiber coupler and beam stabilization system of the GRAVITY interferometer

O. Pfuhl*[a], M. Haug[a], F. Eisenhauer[a], S. Kellner[a], F. Haussmann[a], G. Perrin[b], S. Gillessen[a], C. Straubmeier[c], T. Ott[a], K. Rousselet-Perraut[d], A. Amorim[e], M. Lippa[a], A. Janssen[a], W. Brandner[f], Y. Kok[a], N. Blind[a], L. Burtscher[a], E. Sturm[a], E. Wieprecht[a], M. Schöller[g], J. Weber[a], O. Hans[a], S. Huber[a]

[a]Max-Planck-Institut für extraterrestrische Physik, Giessenbachstraße 1, 85748 Garching, Germany;
[b]LESIA, Observ. de Paris Meudon, place Jules Janssen, 92195 Meudon Cedex, France
[c]I. Physikalisches Institut, Universität zu Köln, Zülpicher Strasse 77, 50937 Köln, Germany
[d]Institut de Planétologie et d'Astrophysique de Grenoble, BP 53, 38041 Grenoble Cedex 9, France
[e]SIM, Fac. de Ciências da Univ. de Lisboa, Campo Grande, Edif. C1, P-1749-016 Lisbon, Portugal
[f]Max-Planck-Institut für Astronomie, Königstuhl 17, 69117 Heidelberg, Germany;
[g]European Southern Observatory, Garching, Germany

## ABSTRACT

We present the installed and fully operational beam stabilization and fiber injection subsystem feeding the 2nd generation VLTI instrument GRAVITY. The interferometer GRAVITY requires an unprecedented stability of the VLTI optical train to achieve micro-arcsecond astrometry. For this purpose, GRAVITY contains four fiber coupler units, one per telescope. Each unit is equipped with actuators to stabilize the telescope beam in terms of tilt and lateral pupil displacement, to rotate the field, to adjust the polarization and to compensate atmospheric piston. A special roof-prism offers the possibility of on-axis as well as off-axis fringe tracking without changing the optical train. We describe the assembly, integration and alignment and the resulting optical quality and performance of the individual units. Finally, we present the closed-loop performance of the tip-tilt and pupil tracking achieved with the final systems in the lab..

## 1. INTRODUCTION

The 4-telescope interferometer GRAVITY[1] is designed to deliver high-resolution imaging of faint objects and narrow-angle precision astrometry in the astronomical K-band (1.9-2.5μm). The instrument combines either the four 1.8m Auxiliary Telescopes (AT) or the 8m Unit Telescopes (UT) of the ESO observatory at Paranal. The large collecting area of the UTs and the capability of fringe-tracking will allow phase-referenced imaging of sources as faint as $m_K$ ~16 at a 3 milli-arcsecond resolution as well as 10 micro-arcsecond precision astrometry. The improvement in angular resolution will be about a factor 15 compared to diffraction-limited imaging on current 10m class telescopes. In its dual-field mode, GRAVITY will provide relative position measurements between two objects with an accuracy of 10μas, i.e. an angle equal to a Euro coin diameter seen at the distance of the moon. Measuring at that accuracy level means observing the universe in motion, since 10 μas/yr correspond to 5m/s at a distance of 100pc (close star-forming regions) and 50 km/s at a distance of 1Mpc (Andromeda galaxy). GRAVITY has been proposed in 2005 as an adaptive optics assisted beam combiner for the second-generation VLTI instrumentation. A detailed description of GRAVITY can be found in Eisenhauer et al. (2011)[1].

Figure 1 tries to depict the working principle of GRAVITY. The 2" VLTI field-of-view (FoV) is corrected by an infrared adaptive optics facility and propagated through the delay lines. Laser beacons injected at the telescopes measure pupil and tip-tilt perturbations in the VLTI. Sensors in the beam combiner instrument track their signal and command actuators in the fiber coupler units. Two objects within the field of view are then coupled into single-mode fibers, adjusted in polarization and in optical path delay and brought to interference in integrated optics chips. The interfered beams are then imaged by two spectrometers. Provided that one of the two objects is bright enough, the instrument can correct atmospheric perturbations by tracking on the bright object. This allows integration on a second faint object within the 2" FoV for timescales significantly longer than the atmospheric coherence time (few 10s). An internal metrology, which measures the differential optical path between the two objects returns the angular separation of the objects at micro-arcsecond precision and allows phase-referenced imaging of resolved objects.

This paper describes the single-mode injection and the beam stabilization of GRAVITY.

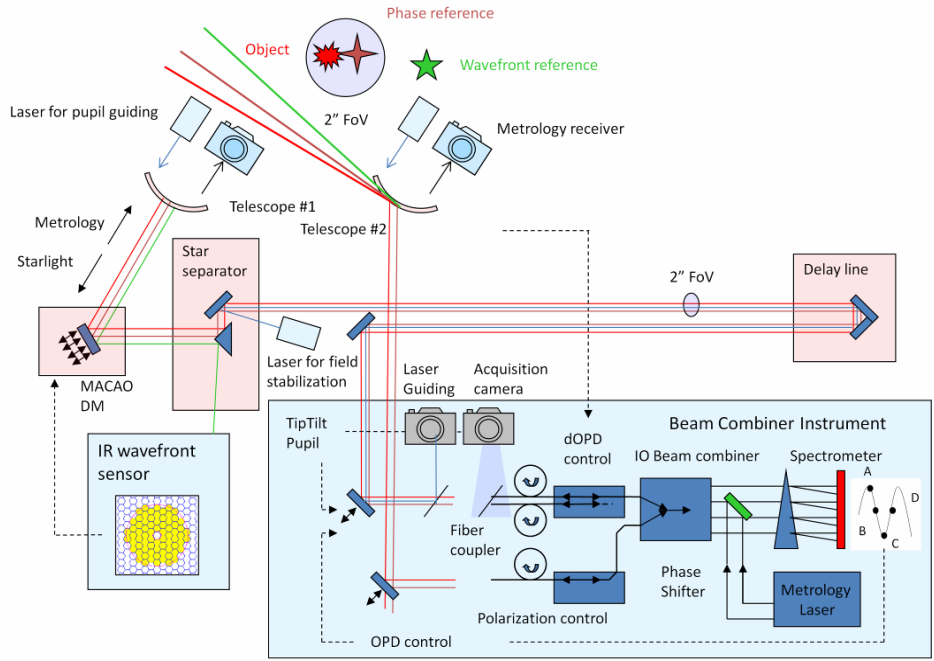

Figure 1: Working principle of the GRAVITY.

## 2. THE FIBER COUPLER

The GRAVITY beam combiner instrument contains four fiber coupler units. The purpose of the units is to pick up and stabilize the beams from the four telescopes, to split the field-of-view (FoV) and to feed the light of two objects into single-mode fibers. Each fiber coupler contains the optics, motors and piezo actuators to rotate field and polarization and to stabilize tip-tilt jitter and pupil wander. Figure 2 shows one of the fiber coupler units. The mechanical structure of the fiber coupler is mounted on the optical bench in the cryostat and kept at a constant temperature of 240K. The temperature is a compromise between the thermal background and the operability of the integrated stepper motors and piezo actuators. One of the stepper motors, equipped with a K-mirror [1] compensates field rotation. Another stepper rotation stage equipped with a half-wave plate [2] allows adjusting and analyzing linear polarization. A piezo-actuated mirror [4] provides fast OPD actuation (fringe-tracking) and tip-tilt adjustment. Pupil wander of the incident beam can be corrected with a second piezo-actuated mirror [5].

The science band is separated from the laser and acquisition bands by a dichroic in the collimated beam [3]. The dichroic reflects the astronomical K-band and transmits the H-band as well as the tip-tilt laser (655nm) and the pupil laser (1200nm). The tip-tilt laser is tracked by a position-sensitive-diode [10] in the laser tracker module [6]. The large bandwidth of the diode allows to track image jitter of up to ~1kHz. Located behind the laser tracker module, the acquisition camera[5],[6] of GRAVITY monitors slow image drifts on sky and at the same time monitors the pupil position.

For calibration purposes, each fiber coupler unit is equipped with an internal retro-reflector that allows mapping the GRAVITY fibers onto the acquisition camera. In this way the fiber positioning can directly be monitored with respect to the object that has to be coupled.

The internal optical path between the integrated optics and the fiber coupler is monitored by a metrology fiber-fed diode located behind the dichroic [3], providing the OPD feedback for the fibered delay lines. Each unit is equipped with a special roof-prism [8] that provides two modes of operation. In the field splitting mode it separates two objects (science and fringe tracking object) within the 2" FoV (UT). In this dual-field mode, the two objects are coupled into two single-mode fibers. After passing the differential delay lines and the polarization control unit, the light of each object coming from the four telescopes interferes in the integrated optics chip. The bright object is used to correct the atmospheric piston jitter and to stabilize the fringes of the second (faint) object. This allows integrating significantly longer than the atmospheric coherence time on the faint object. Instead of a few milli-seconds, the integration time can be as long as minutes, increasing the sensitivity by orders of magnitude. The other mode of the roof-prism acts as a beam splitter, where fringe-tracking and science integration is done on the same object. This particular mode can be chosen for bright objects, where high-resolution spectroscopy is desired.

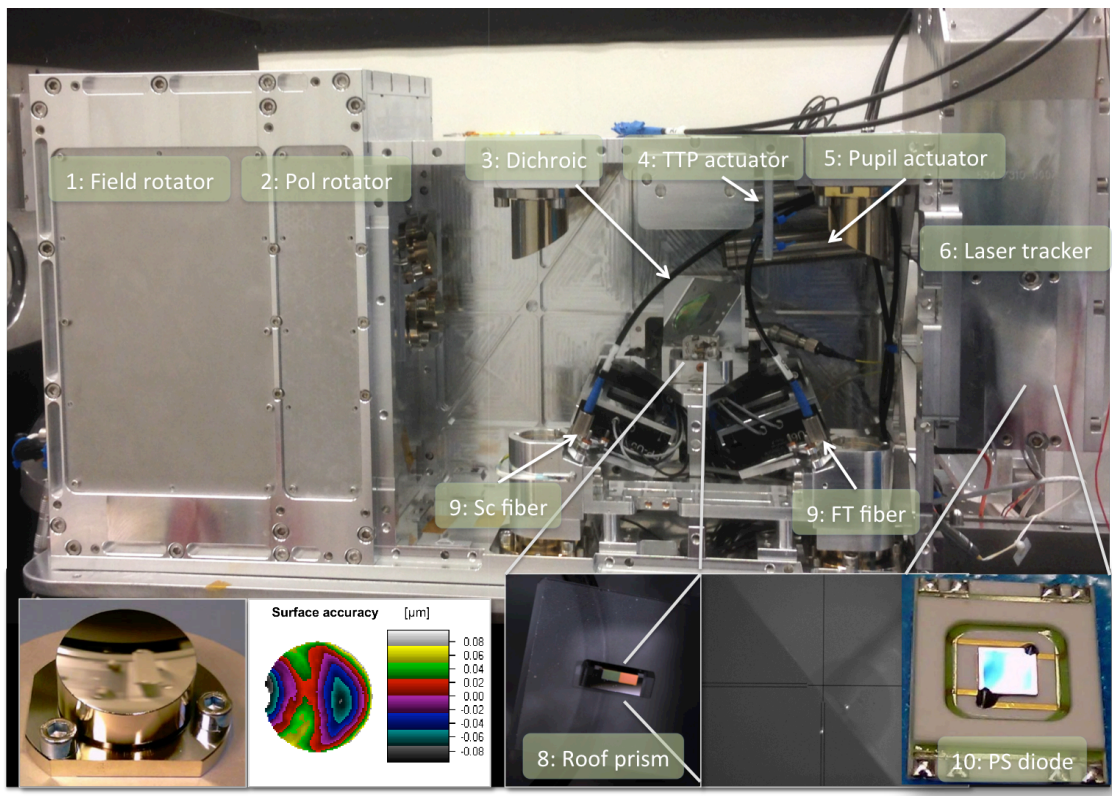

Figure 2: Picture of one fiber coupler unit.

## 2.1 Highlight components

**Diamond-turned mirrors**

Each fiber coupler unit contains 7 off-axis parabolic mirrors made from diamond-turned aluminum. Since aluminum is too soft to be post-polished, a nickel layer is deposited on the substrate. On top of the nickel, a layer of gold ensures high reflectivity in the infrared. The curvature radii are between 104 and 400mm and the off-axis distances are between 28mm and 105mm. The mirror diameters are all about 25mm. Flats at the bottom of the mirrors together with alignment pins define the orientation and alignment of the mirrors. All mirrors turned out to be very good in terms of surface accuracy and surface roughness. The lower inset of Figure 2 shows one of the parabolic mirrors before mounting as well as the measured surface accuracy. Overall the measured surface accuracy of λ/4 P-V at 633 nm (λ/20 RMS) and the roughness of $R_q$ = 1.7nm is excellent and fulfills the requirements.

**Roof-prism**

One of the main requirements for the fiber coupler units is to allow dual-field as well as a single-field operation. In order to avoid optic wheels or other movable parts, the goal was to design an optical element such that the two modes can be offered without additional actuators. Both modes are provided by a special roof-prism provided by Zuend precision optics. The prism substrate is fused silica, which is transparent in the IR. The prism base is $4 \times 1.5mm^2$. The optical active area however is the central $0.3 \times 0.3mm^2$. Each prism is glued on a $10 \times 10mm^2$ N-BK7 glass plate for handling purposes. The top surface of the prism consists of four coated quadrants. Two gold-coated quadrants constitute the typical roof-prism, i.e. two angled mirrors. The tilted mirrors split the field in two half's. The other two quadrants are coated with a beam-splitter- and an anti- reflective (AR) coating. The bottom of the prism is again coated with gold. Thus in the single-field mode, the beam is split 50/50 at the first surface. The transmitted part is reflected at the bottom and exits the prism through the AR coated surface. The angles of the prism are tuned such that the exiting beam travels the same path as the dual-field mode. Changing from dual-field to single-field mode therefore only requires offsetting the internal tip-tilt actuator. The working principle is shown in Figure 3.

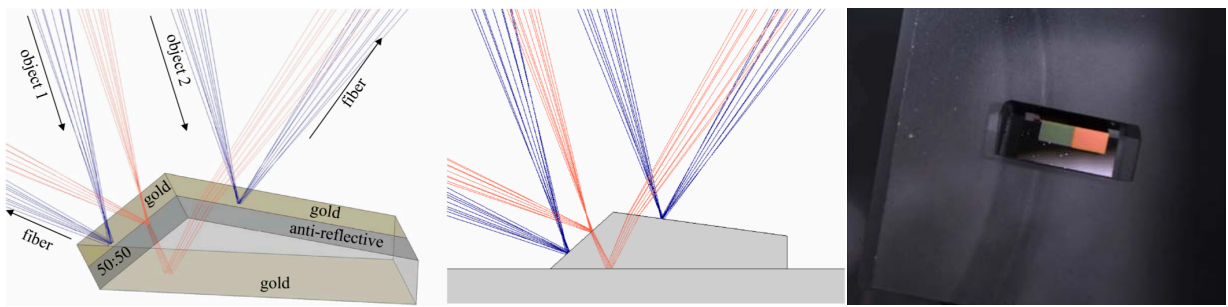

**Figure 3:** (Left) 3D drawing of the roof-prism and the two operation modes (single/dual-field). In the dual-field mode two gold-coated quadrants split the field and reflect the light of two objects (blue rays) in opposite directions. The single- field mode splits the light of one object (red rays) at the first surface. It reflects 50% of the light and transmits the other 50%. The transmitted light is reflected at the bottom of the roof-prism and exits the glass substrate at the opposite side of the prism. (Right) A picture of the roof-prism.

## 2.2 Optical quality

We measured the wave-front error of one fiber coupler unit with a Shack-Hartmann sensor as well as with a FISBA interferometer. The Shack-Hartmann sensor was used to measure the optical path from the entrance aperture to the single-mode fibers, while the interferometer was used to measure the optical path between the entrance aperture and the interface to the laser tracker and the acquisition camera located behind the dichroic (exits as a collimated beam). The results are shown in Figure 4. The overall wave-front error between the entrance down to the fibers is 89nm RMS. At the science wavelength of 2.2μm this corresponds to a wave-front error <λ/24 with a corresponding of Strehl 94%. This is close to the initial goal of 95%.

The wave-front error at the laser tracker interface is 57nm RMS. The wave-front error is lower due to the smaller number of optical elements in the path. Overall the fiber coupler units have an excellent optical quality, which yields an optimum injection efficiency.

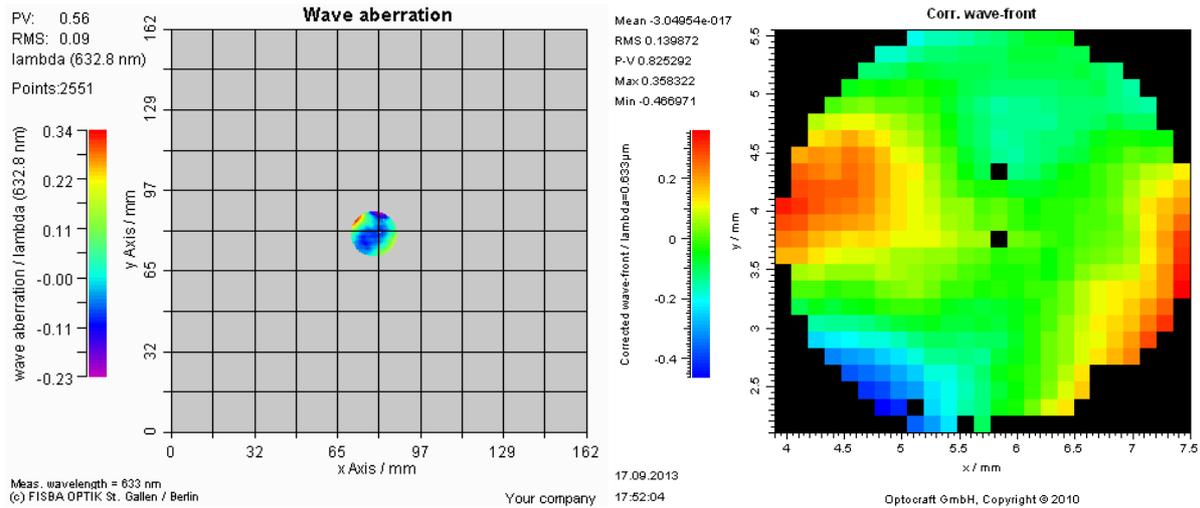

Figure 4: (Left) Wave-front error (tilt subtracted) between the entrance and the interface to the acquisition camera. (Right) Wave-front error between the entrance aperture and the single-mode fibers.

## 3. THROUGHPUT AND ASTROMETRIC ERROR

### 3.1 Effective throughput

Wallner et al have studied the injection loss of a single-mode injection system as function of tilt relative to the fiber mode-field radius[2]. We used their loss curve and converted it to an equivalent angle on sky. The stochastic nature of the atmospheric tip-tilt perturbations leads to an average injection that depends on the overall atmospheric tip-tilt RMS. We used a Monte Carlo simulation with a typical atmospheric tip-tilt power spectrum to derive the average throughput (relative to perturbance free coupling) for various tip-tilt RMS amplitudes. Figure 5 shows the effective coupling efficiency relative to tip-tilt-free coupling as function of tip-tilt RMS.

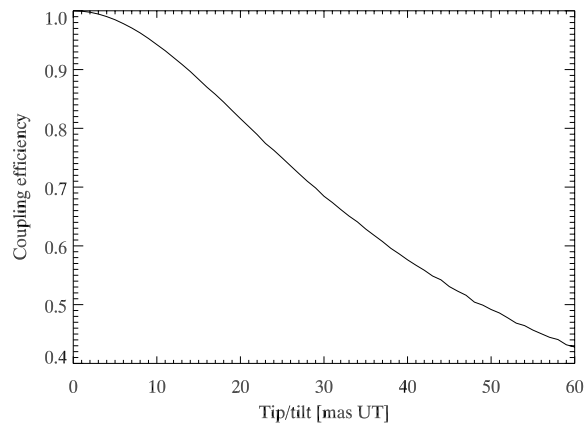

Figure 5: Single-mode coupling efficiency relative to optimum coupling as function of tip-tilt error RMS.

The typical adaptive optics residual tip-tilt is ~10mas. Yet, the dominating tip-tilt source is the VLTI tunnel atmosphere, which contributes a 2-axis tilt RMS of 50mas on sky (UT). The corresponding injection, i.e. throughput loss is thus of the order 50% compared to a perturbation free system.

## 3.2 Astrometric error

The combination of pupil offsets and tip-tilt errors leads to a substantial astrometric error, which can only be mitigated by active tracking of the telescope beams. Figure 6 tries to depict the phase error introduced by pupil and tip-tilt errors. The phase of the beam that is coupled into a single-mode fiber depends on the fiber mode F(u) and the incident electric field E(u), where u is the pupil coordinate. Any mismatch between the pointing of the electric field and the fiber mode (i.e. tip-tilt error) Δα can lead to a phase error in the presence of an additional pupil error Δu. Phase errors immediately translate into angular separation errors. The strict mathematical derivation of that error can be found in Lacour et. al (2014)[7].

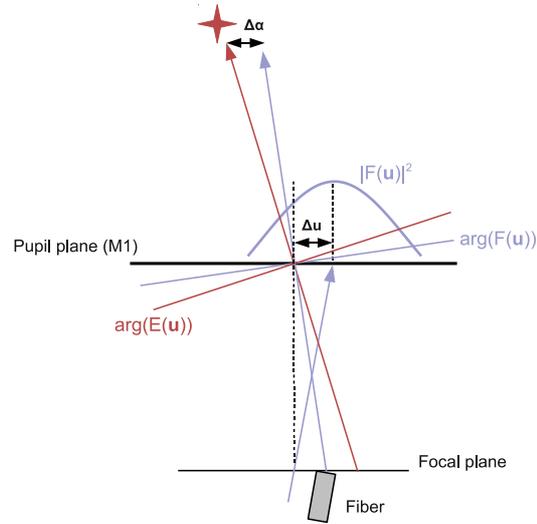

Figure 6: Sketch of the phase error due to combined tip-tilt and pupil error.

In summary the error boils down to:

$$\sigma \approx \frac{\Delta\alpha \cdot \Delta u}{B}$$

, the astrometric error σ is the product of the tip-tilt error Δα times the pupil error Δu normalized by the baseline B. In the absence of those errors or if one of the errors is kept small enough the astrometric error is negligible. Pupil errors however can be up to 10% of the pupil diameter (i.e. 0.8m) and the tip-tilt RMS are of the order 50mas. The resulting astrometric error is thus of the order ~400 μarcsec (for a baseline of 100m), exceeding the envisioned astrometric accuracy by more than an order of magnitude. The mitigation of this combined astrometric error can only be achieved with active beam stabilization.

## 4. BEAM STABILIZATION

In order to achieve the envisioned astrometric precision and the effective throughput, GRAVITY relies on two laser tracking systems. One system stabilizes fast image jitter occurring in the tunnel (tip-tilt perturbation). The other system stabilizes longitudinal and lateral pupil motion, which originates from imperfections in the optical alignment of the telescopes and the delay lines. In fact, the tip-tilt image stabilization consists of two sensors. The tip-tilt laser tracking system corrects fast perturbations in the tunnel using laser beacons, while the acquisition camera guides slowly on a reference star. The pupil guiding uses only the acquisition camera as a sensor, which is sufficient in terms of speed since the pupil drifts only on timescales of minutes. The actuators for the tip-tilt and pupil stabilization are contained in the fiber coupler units (see Figure 2). The joint field and pupil stabilization is required to ensure an optimum coupling efficiency and to reduce the tip-tilt and pupil induced astrometric error.

The atmospheric image jitter originating above the telescope is compensated by the Adaptive Optics (AO) facility of Unit Telescopes[3]. The AO covers the optical path down to the Coude focus (which coincides with the mirror M10 of the star separator). Typical residual jitter after AO correction is ~10mas RMS. The tip-tilt guiding system is launched at the star separator (STS) and propagated to the VLTI lab. In this way the full optical path between the AO and the VLTI lab can be covered.

The pupil guiding uses laser beacons at the telescope spiders. This allows tracking pupil motions between the telescope primary mirror and the VLTI lab.

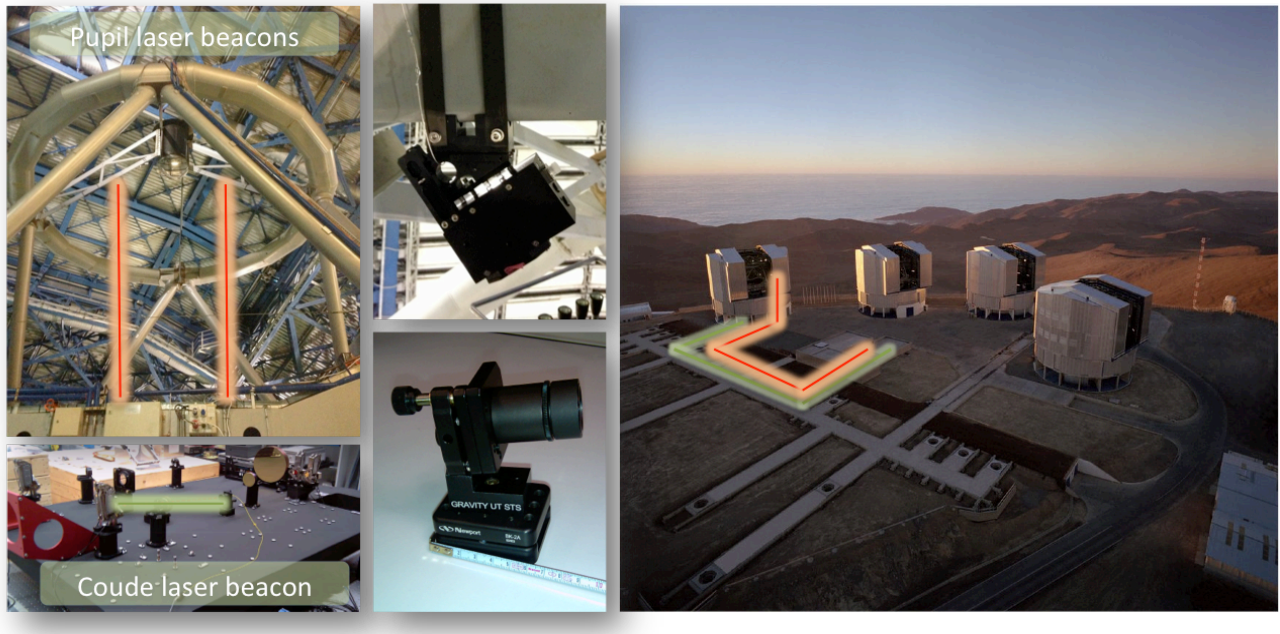

Figure 7: Pictures of the pupil and tip-tilt guiding systems. Four pupil lasers beacons are mounted at the telescope spiders (top left). One pupil laser beacon is shown in the upper middle. The tip-tilt laser beacons are launched in the Coude room (lower right). The corresponding laser launcher is shown in the lower middle. The pupil and tip-tilt laser beacons cover the full path between the telescopes and the VLTI lab (right).

## 5. TIP-TILT TRACKING

### 5.1 Overview

The purpose of the tip-tilt laser tracking is to correct image jitter introduced by the VLTI tunnel atmosphere. The perturbations of the tunnel atmosphere are dominated by the lowest order optical disturbance, i.e. tip-tilt, which accounts for 90% of the wave-front variance. A typical turbulence profile in the tunnel has been established by Gitton & Puech[3] (see Figure 10). One of the main goals of GRAVITY is to observe faint targets. This makes it hard to use existing VLTI infrastructure such as the tip-tilt tracker IRIS, which requires very bright targets for a fast correction. In case of typical GRAVITY targets with $m_K$~11 the required integration time is of the order ~1s, which leaves most of the tip-tilt uncorrected. In order to effectively stabilize fast tip-tilt in the tunnel, a bandwidth of at least several 10Hz is required. This requirement can only be met by a laser-guided system. In order to ensure fast image jitter correction combined with high pointing stability on the stars, we decided for a two-stage tracking system. One fast tracking system (laser tracker), which consists of a bright laser beacon and a position sensitive diode corrects image jitter with a bandwidth of ~100Hz. A second system employs the acquisition camera (star tracker) that monitors the position of a reference star with a bandwidth of ~1Hz. This is ensures an optimum pointing stability and cancels any differential drifts between the beacon and the actual object on sky. The star tracker works as a setpoint control for the laser tracker. When the star position needs to be adjusted, the system sends a new relative setpoint to the laser tracker. This causes the controller of the laser tracker to offset the actuator to the new position. The corresponding setpoint control loop is depicted in Figure 8.

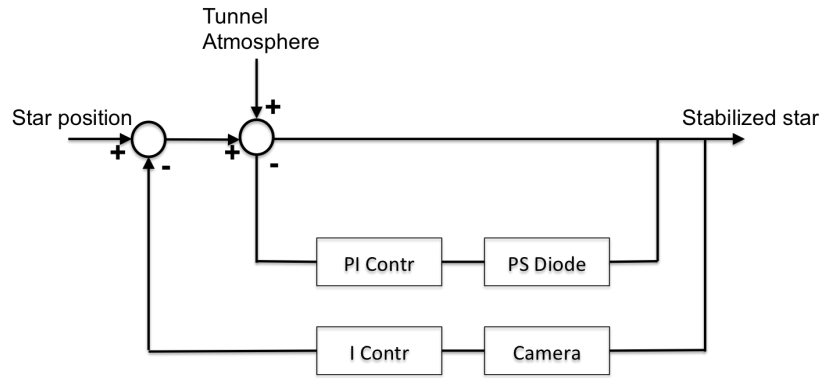

**Figure 8: Setpoint control of the tip-tilt guiding system. The slow star tracker sets the setpoint for the fast laser tracker.**

## 5.2 Hardware

**Tip-tilt laser beacons**

Each Coude focus will be equipped with one tip-tilt laser beacon (depicted in Figure 7). The beacon will be launched at the star separator in a field plane. In order to inject the laser should not vignetting or adding any optical elements to the VLTI optical train, we decided to focus the laser onto the field mirror star separator. This creates a bright scattering spot in the respective plane. The laser itself is launched from below the optical axis with an angle of incidence of 5.8°. In this way, an artificial guide star is introduced in the FoV. Tracking on this artificial star therefore allows stabilizing the image motion. The incidence angle is as small as possible since the scattering efficiency strongly depends on the angle between the specular reflection and the scattering direction. Nearly all the scattered power is radiated away in a small cone (< 10°) around the specular reflection. In our injection scheme, the specular reflection itself will be completely baffled within the star separator enclosure. The optical design of the tip/tilt laser beacon is such that it contains only off-the shelf optical components. The laser source is a commercial fiber-pigtailed laser diode with a wavelength of 658 nm and a nominal output power of 60mW. A constant current laser driver housed in the STS LCU provides the power for the laser. The laser can be switched on and off, via a digital I/O.

**Position-Sensitive-Diode**

The tip-tilt beacon is imaged onto a position sensitive diode (PSD). A PSD consists of a uniform resistive layer formed at one or both surfaces of a high-resistivity semiconductor substrate, and a set of electrodes formed on the ends of the resistive layer for extracting position signals. The active area, which is also a resistive layer, has a PN junction that generates a photocurrent if light hits the surface. The photocurrent that is measured at a particular electrode is inverse proportional to the distance between the light spot and the electrode due to the increasing resistance. By comparing the photocurrents at the output electrodes, the spot position can be retrieved. The PSD used in the laser tracker is a silicon duo-lateral PSD from Pacific Silicon Sensors. An extremely low-noise amplifier board developed by MPE amplifies the four output voltages. The amplifier works at a very high amplification with a 3 dB bandwidth of 820Hz. The signal is then recorded by a real-time LCU, where the resulting X-Y position is calculated.
Taking into account the scattering efficiency, the VLTI transmission and the GRAVITY instrument transmission, only about 1μW of the initial 60mW laser will reach the PSD.
We measured the position accuracy of the PSD and compared it to theoretical noise calculations (see Figure 9). It turns out that at 1μW incident power, the position uncertainty is about $dL/L = 0.1\%$. At this light level, material inhomogeneities dominate the noise. The same performance can be reached even if the light level is a factor ten lower than anticipated. That gives us a comfortable margin for the operation of the system.

The position uncertainty of dL/L = 0.1% translates into a position resolution on sky by multiplication with the equivalent FoV of 2.3", resulting in a sensor noise of ~2.3mas RMS (UT).

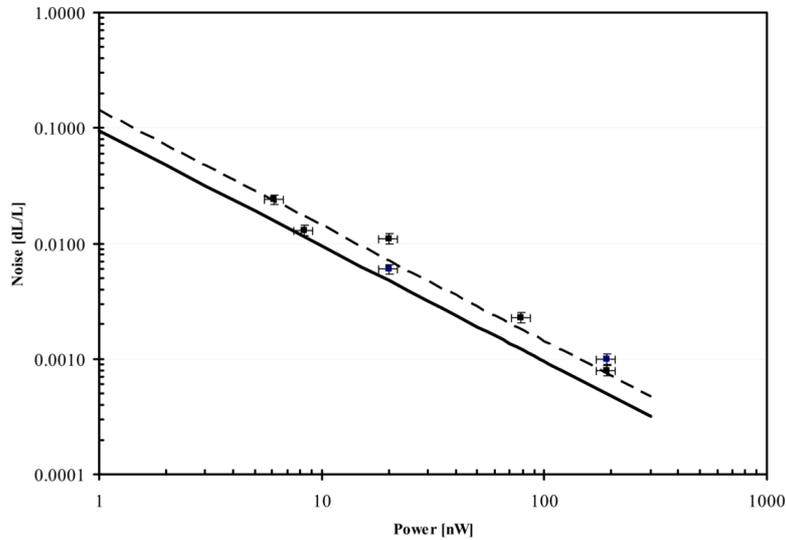

**Figure 9: RMS Position noise [dL/L=4mm] as function of the incident light level of the diode in use. The data-points mark the measurements. The theoretical noise is shown in black (solid) and the noise scaled to the measurement (dashed).**

**Actuator**

The tip-tilt/piston actuator is a Physik Instrumente S-325.3SL piezo actuator (see Figure 2). Each piezo unit is equipped with an internal feedback sensor. Given that the actuator operates at 240K and in vacuum, strain-gauge sensors replaced the normal capacitative sensors. Several units of this type have already been tested and characterized at MPE. The full travel range of the actuator is 30μm, or almost 60μm of optical path at a ~1nm resolution. In terms of tilt, the full stroke is 4mrad or 8mrad optical tilt at 0.2μrad resolution, corresponding to a stroke of 3.7" and 0.09mas resolution on sky. Therefore the full FoV of the VLTI can be accessed, while leaving some margin for alignment corrections. The 3dB bandwidth of the piezos is about 220Hz.

## 5.3 Measured closed-loop performance

In order to test the laser tracking closed-loop performance, we applied a perturbation similar to the tunnel atmosphere to system. The perturbation was modeled as a first-order filter with a cutoff frequency of 1Hz. The resulting power spectrum and the comparison with the model tunnel atmosphere[1] can be seen in Figure 10. The input perturbation was scaled such that the total RMS was an equivalent to 36 mas on sky (typical tunnel atmosphere, 1-axis). After closed loop-correction (Figure 10), the tip-tilt RMS was reduced to 3.0 mas. The performance is close to the sensor noise limit, which is of the order 2 mas. This means that the system is able to reduce the tunnel perturbations by about a factor 10. Naturally the system can only correct perturbations occurring in the tunnel. Tip-tilt residuals from the AO system cannot be corrected. According to simulations performed by Clenet & Gendron[4], about 10mas RMS tip-tilt are not corrected by the AO facility. Thus the laser guiding system reduces the tunnel tip-tilt to a level, that the AO residuals are limiting the injection efficiency.

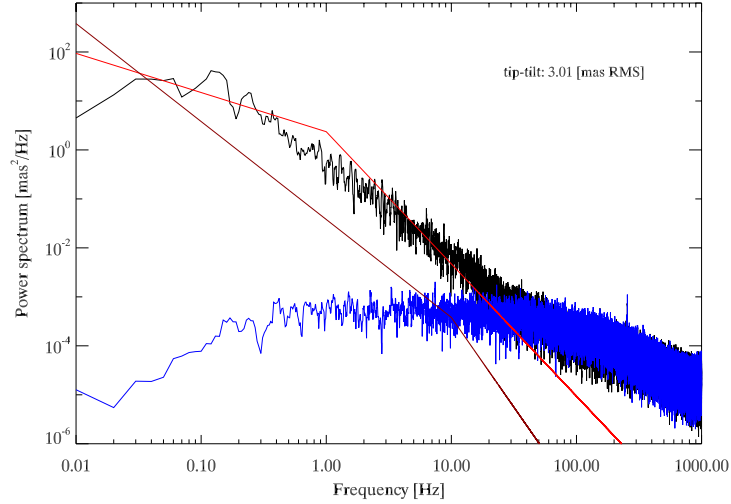

Figure 10: Closed-loop measurement of the tip-tilt laser guiding. The input perturbation is shown in black. It closely resembles the typical VLTI tunnel atmosphere power-spectrum (red). An alternative atmosphere model is shown in dark red. The closed-loop response is shown in blue.

Without tip-tilt correction the typical 36 mas RMS tip-tilt (2-axis: 50 mas RMS) reduce the injection efficiency to about 50% compared to the injection without tip-tilt perturbation (see Figure 5). The laser guiding system with a residual tip-tilt of 10 mas (limited by the AO) reaches an injection efficiency of 94%. This means that the effective throughput of GRAVITY will be increased by almost a factor two due to the active guiding in the tunnel.

## 6. PUPIL TRACKING

The VLTI optical train suffers from significant pupil movements during the course of a night. This is partly due to temperature stratifications but mainly due to delay line wobble. Depending on the delay line position and the time between re-alignments the pupil runout can be as large as 20%[3]. Given that one major astrometric error source originates from the combination of pupil shift and tilt error[7], also the pupil has to be actively stabilized. However, the nature of the pupil error is different than the tip-tilt error. While the field jitter is introduced due to the random fluctuations of the tunnel atmosphere, the pupil moves due to mechanical defects and temperature gradients. Changing telescope pointings and delay line positions, leads to pupil wander in the lab. The largest lateral pupil displacement can be attributed to the wobble of the delay line carriage. In fact, the displacement shows a periodicity connected to the circumference of the carriage wheels. Another pupil misalignment is related to the model accuracy of the variable curvature mirror (VCM). Since the delay lines move during an observation, the VCM has to constantly adjust the longitudinal pupil, to keep it at the nominal position in the VLTI lab. Changing the curvature of a field mirror according to a blind pointing model does this adjustment. The pointing model limitations and the wobble lead to a pupil wander in the tunnel of up to 4% of the pupil diameter (3mm shift of the 80mm beam) in lateral and 1m in the 80mm beam in longitudinal direction (Gitton & Puech[3]). It is convenient to treat the longitudinal pupil shift as a field dependent lateral pupil position. The fiducial 2" FoV on sky of the 8m UT translates into a field angle in the 80mm beam of 2" · 8000/80 = 3.3'. Therefore the effective lateral pupil error due to 1m longitudinal pupil shift is 3.3' ·1m ~ 1mm or an equivalent of 1.3% of the pupil diameter. Thus the error is of the same order as the delay line wobble, and both terms have to be corrected. The astrometric phase error is a product of the residual tip-tilt and pupil errors (we refer to Lacour et al.[7] for further reading). Therefore the required pupil stability depends on the achievable tip/tilt correction. Based on the estimated residual tip-tilt of 10mas RMS, the residual pupil shift has to be < 0.5% of the pupil, i.e. < 40mm at M1.

## 6.1 Hardware

**Pupil beacons**

Each telescope will be equipped with four pupil-guiding beacons installed at the telescope spiders (see Figure 7). The reason to use four beacons is that the pupil rotates during an observation and that the center of the pupil cannot be equipped with a beacon (occupied by M2). Four beacons placed symmetrically around M2 allow tracking the beacon barycenter, which is not affected by pupil rotation. The four beacons are re-imaged within the beam combiner instrument by the acquisition camera pupil tracker[5],[6]. We chose a laser wavelength of 1200nm since both AO wave-front sensors (infrared and visible) are blind at that wavelength. About 1nW of laser power in case of the UTs and 0.05nW in case of the ATs is sufficient to provide enough signal-to-noise on the acquisition camera. The pupil launcher opto-mechanical design is identical for the UT and the AT case. Only the mount that fixes the launcher box to the telescope spiders is different. For the laser source, a 5mW laser diode from Frankfurt Laser Company is coupled into a multi-mode fiber. Three fiber splitters distribute the light onto four output fibers. All splitters and fiber collimators in use are off-the shelf products with FC connectors. To reduce the laser power from mW to nW, the output fibers are equipped with fiber attenuators. The diode, the coupling optics and the fiber splitters are located in an electronic box that will be mounted at the telescope. The box provides four FC connectors to feed the relay fibers that are routed along the telescope spiders up to the launch boxes.

**Sensor**

The pupil tracker of the acquisition camera[5] provides the feedback for the pupil stabilization loop. The optical and mechanical development of the acquisition camera is the work-package of the University of Lisbon. However, since the pupil tracker is the sensor for the pupil guiding system, its working principle and design is briefly described here. For a more detailed description of the unit, we refer to Amorim et al.[6].

We developed a pupil-tracking concept that is able to correct lateral and longitudinal pupil drifts using laser beacons installed at the telescope spiders (close to the pupil plane). The concept uses a lenslet array, similar to a Shack-Hartmann wavefront sensor, where the longitudinal pupil error translates into a focus error. The lenslet, placed in the conjugate of the pupil plane (a field plane), provides multiple images of the beacon. In this way, lateral and longitudinal pupil motion can be detected simultaneously. While lateral pupil motion leads to a common shift of the images, longitudinal motion leads to a contraction or expansion of the spot pattern, depending on the direction. The concept is illustrated in Figure 11.

In the actual implementation, we decided to use a 2 × 2 lenslet array (see Figure 12, right). Since we use four beacons, the resulting detector image contains four times four spots (see Figure 12, left). The lateral pupil shift can be inferred from a common lateral shift of all four barycenters, while a longitudinal shift can be deduced from the barycenter distance across the pupil images.

**Actuator**

The pupil actuator is a Physik Instrumente S-330.8SL piezo actuator (see Figure 2) with a 10mrad tilt stroke (20mrad optical tilt). Located at the focus of a parabola with 200mm focal length, the actuator tilt can be converted into a pupil adjustment stroke of $20 \cdot 10^{-3} \cdot 200mm = 4mm$. Given that the pupil diameter in the VLTI lab is 18mm, the relative pupil stroke is 22%. The actuator is required to compensate the pupil wander that occurs between the telescope and the lab. The main causes of the pupil wander are the delay line wobble and the telescope run-out, which can move the pupil by ~few mm. However the pupil wander happens on timescales of minutes and therefore the actuator response of only 10Hz is sufficient.

**Controller**

The controllers for the longitudinal and the lateral pupil control are in both cases proportional-integral controllers. The sensor in both cases is the acquisition camera, which runs at a frame rate of 1Hz. The actuators however, are different. The lateral pupil control uses a piezo in the fiber coupler (see Section 2). The longitudinal pupil position has to be adjusted by the variable curvature mirror (VCM) of the main delay line. We modeled the controller as a normal integral controller with sensor integration time of 1s, a computational delay of 0.5s and an actuator with a 10Hz bandwidth.

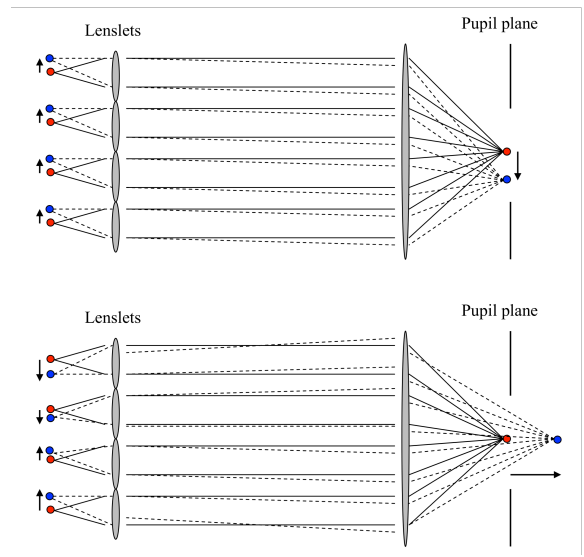

Figure 11: Pupil tracking concept. The shift pattern of the reimaged pupil spots allows measuring lateral and longitudinal pupil wander.

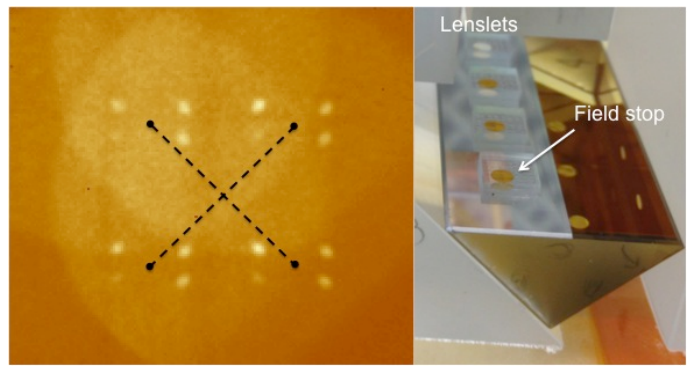

Figure 12: Detector image of the four laser beacons imaged by four lenslets (left). The barycenter of the four beacons imaged by the individual lenslets is indicated in black. Prism with the field stop and the rectangular lenslet glued on top (right).

### 6.2 Simulated closed-loop performance

Data obtained 2010 during a dedicated pupil motion measurement campaign in Paranal allowed us to apply the simulated controller to the data. The data contains the measured pupil wander due to delay line motion over 20m of optical path at a speed of 10mm/s. This represents rather extreme observing conditions with quickly changing optical path. Therefore it is well suited to estimate the system performance. The measured pupil motion and the simulated residual motion after closed-loop correction is shown in Figure 13. After closed-loop correction the residual lateral pupil jitter is reduced from 2.5 mm to 0.07mm RMS (tunnel). This corresponds to < 0.1% RMS pupil diameter. The longitudinal pupil correction leaves a residual jitter of < 0.1m RMS in the tunnel. This corresponds to a lateral pupil shift over 2" FoV of < 0.13% pupil diameter. In other words, the pupil guiding system reduces the lateral pupil motion by a factor 40 and the longitudinal motion by a factor 10. The total error is 0.16% of the pupil diameter (12mm @ M1). The resulting pupil error terms relevant for the astrometric error are of the same magnitude. Overall the system easily fulfills the required pupil stability for GRAVITY of residual lateral pupil motion < 0.5%.

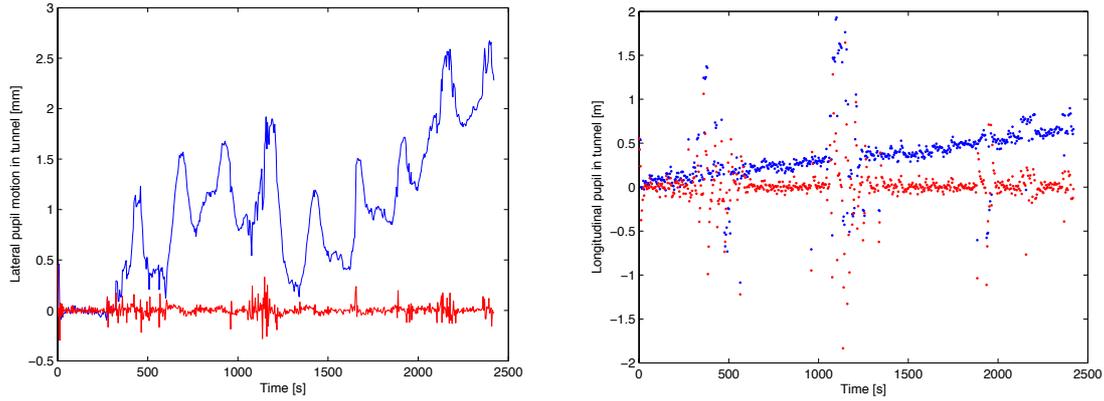

**Figure 13:** Measured pupil wander (lateral ->right, longitudinal ->left) during 20m delay line motion with 10 mm/s speed (blue). The distinct peaks (left) correspond to the circumference of the carriage wheel. The residual wander after closed-loop correction is shown in red.

**Impact on astrometric error**

The combination of pupil and tip-tilt tracking reduces the tip-tilt from 50mas to a residual 10mas RMS and the pupil wander from ~4% to a residual 0.16% (12mm) RMS. For the fiducial baseline of 100m the astrometric error is reduced from 400 µarcsec to 1 µarcsec (see Section 3.2). This means that the ambitious goal of 10 µarcsec astrometry with GRAVITY can be reached with the here presented beam stabilization system.

## 7.  CONCLUSIONS

- The fiber coupler units that have been developed built and tested ensure an efficient single-mode coupling of the telescope beams into the beam combiner instrument. Each unit provides the necessary functions to rotate the field on sky, to adjust the linear polarization orientation, to fringe-track, and to adjust the field and the pupil positions of one telescope beam. It splits the beam with a dichroic and feeds the science band (1.95-2.45µm) into single-mode fibers, while the acquisition band (1.5-1.8µm) as well as the guiding lasers are propagated to the acquisition camera and the laser tracker.

- A corner stone of the optics is the roof-prism. Although only 1.5×4mm$^2$ in size, the prism layout allows to either split the FoV in the dual-field operation of GRAVITY or to use it as a beam splitter in the single-field operation

- The measured wave-front error of the fiber coupler system is ~89nm RMS. This results in an excellent Strehl in the K-band of 94%.

- The combination of the acquisition camera slow image guiding and the fast tip-tilt laser tracking allows correcting most of the tunnel atmosphere. The system has been measured to reduce tip-tilt perturbations by a factor ten. This essentially cancels the tunnel contribution to a residual 3mas RMS. This only leaves the AO tip-tilt residuals of 10mas, which cannot be corrected.

- The compensation of the tunnel tip/tilt significantly increases the coupling efficiency. In the current state, i.e. without beam stabilization, on average only about 50% of the light is coupled into the fibers. The fast tip-tilt tracking increase the effective coupling to 94% (limited by the AO performance). This corresponds to an increase of the effective throughput by a factor two.

- The pupil guiding system can simultaneously correct VLTI lateral and longitudinal pupil wander. Four 1200nm laser beacons, installed at the telescope spiders, are imaged by a lenslet array on the acquisition camera. The resulting spot pattern allows retrieving the pupil position along and perpendicular to the optical axis.

- The control-loop model shows, that the pupil guiding system can reduce lateral pupil wander from 4% to about 0.1% of the pupil diameter and the longitudinal pupil from 1m to about 0.1m position uncertainty in the 80mm beam. The overall pupil error is <0.16% (12mm @ M1).

- The astrometric error due to beam instabilities is reduced from 400 μarcsec for an un-stabilized beam to ~1 μarcsec if the beam is actively stabilized with our system. This allows reaching the ambitious goal of 10 μarcsec astrometry with GRAVITY.